\documentclass[12pt,a4paper]{article}
\usepackage{graphicx}
\usepackage{times}
\title{\bf First-light instrument for the 3.6m DOT: 4Kx4K CCD Imager}
\author{S. B. Pandey$^1$\thanks{shashi@aries.res.in}, R. K. S. Yadav$^1$, Nandish Nanjappa$^1$, \\
Shobhit Yadav$^1$, B. Krishna Reddey$^1$, Sanjeet Sahu$^1$ and R. Srinivasan$^2$\\
\vspace{1cm}\\
\normalsize $^1$ ARIES, Manora Peak, Nainital, UK, 263001\\
\normalsize $^2$ Vemana Institute of Technology, Mahayogi Vemana Road, Bangalore, 560034}
\date{\mbox{}}
\begin{document}
\maketitle
\pagestyle{empty}
%
%
\def\bull{\vrule height .9ex width .8ex depth -.1ex}
\makeatletter
\def\ps@plain{\let\@mkboth\gobbletwo
\def\@oddhead{}\def\@oddfoot{\hfil\scriptsize\bull\quad
"First Belgo-Indian Network for Astronomy \& astrophysics (BINA) workshop'', held in Nainital (India), 15-18 November 2016 \quad\bull}%
\def\@evenhead{}\let\@evenfoot\@oddfoot}
\makeatother
%
%
\def\beginrefer{\section*{References}%
\begin{quotation}\mbox{}\par}
\def\refer#1\par{{\setlength{\parindent}{-\leftmargin}\indent#1\par}}
\def\endrefer{\end{quotation}}
%
%
	{\noindent\small{\bf Abstract:} 
	As a part of in-house instrument developmental activity at ARIES, the 4K$\times$4K CCD Imager is
	designed and developed as a first-light instrument for the axial port of the 3.6m
	DOT. The f/9 beam of the telescope having a plate-scale of $\sim$ 6.4 arc-sec/mm is utilized to
	conduct deeper photometry within the central 10 arc-min field of view. The pixel size of
	the blue-enhanced liquid Nitrogen cooled STA4150 4K$\times$4K CCD chip is 15 micron,
	with options to select gain and speed values to utilize the dynamic range.
	Using the Imager, it is planned to image the central $\sim$ 6.5$\times$6.5 arc-min$^2$ field of view 
	of the telescope
	for various science goals by getting deeper images in several broad-band filters for point sources
	and objects with low surface brightness. The fully assembled Imager along with automated
	filter wheels having Bessel $UBVRI$ and SDSS $ugriz$ filters was tested in late 2015 at the
	axial port of the 3.6m DOT. This instrument was finally mounted at the axial port of the
	3.6m DOT on 30$^{th}$ March 2016 when the telescope was technically activated jointly by the Prime
	Ministers of India and Belgium. This instrument is expected to serve as a general purpose
	multi-band deep imaging instrument for variety of science goals including studies of cosmic
	transients, active galaxies, star clusters and optical monitoring of X-ray sources discovered by
	the newly launched Indian space-mission called ASTROSAT and follow-up of radio bright objects
	discovered by GMRT.
	}
	%
	%
	\section{Introduction}
	Devasthal (Latitude: 29 deg 23 min North, Longitude: 79 deg 41 min East, Altitude: 2540 m)
	, a new observing station of Aryabhatta Research Institute of Observational Sciences (ARIES)
	is well suited for astronomical observations at optical and near-infrared wavelengths in terms
	of seeing (10 \% of the nights, FWHM seeing is 0.7 arcsec), atmospheric stability
	($\sim$ 210 spectroscopic and 160 photometric nights per year) and having excellent logistics
	for any high altitude observational facility (Pant et al. 1999, Stalin et al. 2001). Research 
	activities in many Galactic and extra-galactic scientific areas have been conducted at ARIES
	during last few decades (Pandey \& Mahra 1986, Stalin et al. 2004, Joshi et al. 2006, Yadav et al. 2008) 
	using meter-class telescopes. ARIES Nainital has longitudinal advantage
	(as it lies in the middle of the 180$^{\circ}$ wide belt between Canary Islands 
	($\sim 20^{\circ}$ W) and Eastern Australia ($\sim 160^{\circ}$ E) for observations of 
	transients and time-critical astronomical events like Gamma-ray bursts (GRBs) 
	and Supernovae (SNe) and micro-lensing events (Sagar \& Pandey 2012, Pandey 2006, Joshi et al. 2005).

	\noindent
	Recently installed Devasthal Optical Telescope (DOT) has a f/2 primary 
	mirror of diameter 3.6-m and the secondary mirror has a diameter of 0.9 m
	with an effective focal ratio of f/9, Ritchey-Chretien configuration and 
	Alt-Azimuth mounting. The field of view of the DOT is 10 arcmin without corrector
	at the axial port whereas 30 arcmin unvignetted field for axial port and 10 arcmin 
	for the two side ports. The tracking accuracy of the 3.6m DOT is smaller than 0.1 arcsec 
	RMS for less than one minute in open loop and smaller than
	0.1 arc sec RMS for about one hour in closed loop mode (see Sagar et al.
	2012 for more details and other articles in this proceeding)
	A cylindrical space of minimum 2.5 meter below the focal plane for axial 
	port and approximately 3.0 meter diameter around optical axis will be 
	available for the instrument envelope. Several first-generation back-end
	instruments were proposed for the 3.6m DOT for broad-band imaging and 
	spectroscopy covering 350-3600 nm wavelength range i.g. 
	(1) 4K$\times$4K CCD optical general purpose Imager for deeper photometric 
	observations, (2) TIFR near-infrared imaging camera (TIRCAM2), 
	(3) ARIES Devasthal Faint Object Spectrograph and Camera (ADFOSC), 
	(4) high resolution Echelle spectrograph, (5) a TIFR-ARIES near-infrared spectrometer 
	(TANSPEC) and (6) multi- integral field unit optical spectrograph (DOTIFS). 

	\noindent
	In this article, we describe about design development of the first-light 
	instrument called 4K$\times$4K CCD optical Imager along with some of the 
	proposed science goals. Also, we briefly describe about results of the characterization 
	of the instrument and some of the preliminary science results based on the 
	multi-band imaging performed with the CCD Imager mounted at the axial port of the
	3.6m DOT.

	\section{Science Objectives}\label{sec:intro}

	Deeper imaging of Galactic and extra-galactic point sources and objects with 
	low surface brightness could be performed using the 4K$\times$4K CCD Imager 
	in several broad-band filters (set of Bessel $UBVRI$ and SDSS $ugriz$ filters) 
	at the axial port of the 3.6m telescope. Broad areas of various science 
	objectives are briefly described in the following sub-sections.

	\subsection{Star clusters as a tool for stellar evolution:}
	A group of gravitationally bound stars formed nearly at the same time from the molecular
	cloud are known as star clusters. They are therefore located at the same distance and also
	considered to have same primordial chemical composition. Consequently, HR-diagram of a star
	cluster reflects the evolutionary dispersion of approximately equally old stars of different
	masses, after a lapse of time specified by the age of the cluster. Open clusters and globular
	clusters therefore provide a wealth of information against which to test the theory of
	stellar evolution. Comparison between the theoretical and observed HR (Hurtzsprung Russell)
	diagrams leads to the determination of the age and chemical composition of stars in the cluster.
	(Sagar et al. 1986, Friel 1995, Yadav et al. 2008, Saurabh et al. 2012).
	\noindent
	Using multi-band data it has been observed that most of the metal rich globular clusters in the 
	Galaxy are concentrated towards the Bulge and are severely affected by interstellar absorption
	(Sagar et al. 1999, Alonso-Garcia, J. et al. 2012) except in the Baade's window (Stanek 1996). 
	Consequently, only a few have been observed
	down to the magnitude of the turn-off point in the CMD, and hence their ages are unknown.
	Use of infrared fluxes along with optical should allow a more precise conversion from observed to
	theoretical parameters and hence improved isochrone fitting to be made. It should thus be
	possible to determine the ages of the metal rich clusters with high precision and this will allow 
	a comparison to be made with the ages of the metal poor clusters (William \& Rene 1979, Bekki \& Forbes
	2006, Kalirai et al. 2008).

	\subsection{PMS stars and star formation in star clusters:}
	The presence of dust and gas within young star clusters and in their immediate vicinity is well
	established. Therefore their properties and the role in the evolution of a cluster are crucial
	in understanding star formation as well as cluster formation mechanism. The knowledge of
	extinction properties in these systems is necessary to determine accurate physical parameters
	using optical data along with data taken at near-infrared wavelengths. Young open clusters (YSCs; age 
	$<$ 10 Myr) still embedded in the parent nebulous regions, which are neither evolved
	into stellar remnants nor escaped the gravitational grasp of the molecular cloud and/or dense stellar
	cluster, present an unique laboratory for understanding the process of star formation, its history, the
	early evolution of stars over a wide mass range and the nature of interaction of young stars with
	their surrounding interstellar medium. These regions contain stars of high mass as well as low mass
	pre-main sequence (PMS) stars and the characteristics of their physical properties along with the
	distribution of ionized gas and molecular cloud under the influence of high mass stars can give us clues
	about the physical processes that govern their formation. Broad mass range of the cluster members
	can also be used to quantify the relative numbers of stars in different mass bins by using optical HR diagrams
	and to find systematic variation of the IMF with different star-forming conditions. Identifying systematic 
	variations of star formation would allow us to understand the physics involved in assembling each of the 
	mass ranges and thus to probe early cosmological events (Kroupa 2002).
	IMF of young clusters in the mass range about $\sim$ 30 − 0.6M$_\odot$ can be represented by power law with
	a break around 1-2M$_\odot$. It is found that IMF varies from one cluster region to another (Saurabh 2012).
	The evolutionary sequences and the nature of these PMS 
	stars are still not understood. The imaging of such objects would be able to determine their 
	age correctly and will also be helpful to constrain the mass-function towards lower mass side
	(Pandey et al. 1997, Pandey et al. 2008).

	\subsection{Study of GRB afterglows and Supernovae:}
	Recent studies suggest that the observed burst of Gamma-rays (GRBs) and/or core-collapse supernovae (CCSNe) 
	occur due to the conversion of the kinetic energy in relativistic out-flowing ejection from the progenitor 
	(i.e. central engine of the GRB or so called fireball) to non-thermal gamma-ray/X-ray radiation via internal 
	shocks. The subsequent interaction of the ejected material with the surrounding medium through 
	external shocks produce afterglows visible in all bands from X-rays to radio wavelengths which could be 
	predominantly thermal radiation in case of CCSNe. 
	Afterglow observations of GRBs/CCSNe are of crucial importance in understanding this highly energetic, 
	cosmological, ultra-relativistic phenomenon. Observations of optical afterglows play unique 
	role in understanding the nature of their progenitor, environment and energetics (Pandey et al. 2003,
	Sagar \& Pandey 2012)). 
	Usually the afterglow of GRBs observed at $\sim$ 16-18 mag and fainter (Kann et al. 2011). In few cases, 
	afterglows are very faint i.e. 23 mag and fainter. Photometric measurements are also necessary 
	to know more about the surroundings of the progenitors, nature of the dark GRBs and the host 
	galaxies (Kumar \& Zhang 2015). The correlation of some of the GRBs with SNe could also be probed using photometric 
	observations at late times i.e. $\sim$ 15(1+z) days, where z is the measured red-shift of the burst. 
	Optical afterglows of short-duration GRBs, Kilonovae and other 
	interesting energetic transients and newly discovered various types of Supernovae 
	are also aimed to be monitored with the 4K$\times$4K CCD Imager at the 3.6m DOT (Pandey 2013).  

	\subsection{Galaxies and Optical variability of powerful active galactic nuclei:}
	Much progress in extra-galactic research and cosmology rests on a detailed knowledge of the 
	properties of nearby galaxies. Deep CCD photometry gives direct evidence of various star-forming 
	regions and dust lanes in the galaxies and also incidence of merger or interaction of galaxies. 
	There is still a large sample of galaxies to be studied. Wolf-Rayet galaxies are another 
	important class of objects, both in terms of understanding massive star evolution and star burst 
	phenomenon. Also, optical intensity fluctuations of radio-loud quasars on hour-like time scale 
	is now a well established phenomenon. In contrast, it is not at all clear if similar intra-night 
	variations are also exhibited by radio-quiet quasars which comprise nearly 80\% of the quasar 
	population. Such observations can provide valuable clues to the question of radio dichotomy of 
	quasars, currently among the most outstanding issues in extra-galactic astrophysics. 
	The significance of intra-night variability studies lies in the fact that they enable a probe 
	of the innermost nuclear cores of active galaxies, on the scale of micro-arcseconds which are 
	totally beyond the reach of any imaging techniques in use currently. Multi-wavelength study of 
	rapid intensity variations of active galactic nuclei (AGNs) thus provides a uniquely powerful tool for 
	investigating the process occurring in the vicinity of their central engines. 
	Given the widespread perception that most radio-quiet objects 
	lack relativistic jets, the likely sources of these small fluctuations are the accretion disks 
	that almost certainly exist around central super massive black holes. However, no consensus 
	has been reached about how similar the intra-night optical variability characteristics are 
	for radio-quiet and radio-loud quasars (Stalin et al. 2004, Fabian 2012). 

	\subsection{Studies of EUV-bright and soft X-ray sources:}
	So far only about 35\% of ROSAT sources have been identified with known optical objects. 
	The optical identification content of these sources is equally deficient (Pasquini \& Belloni 1991). 
	Extrapolating from the 
	number and types of sources that have been identified so far, it is expected that about 50\% of 
	these sources will be AGNs or quasars, 35\% will be stars, 10\% will 
	be normal galaxies and clusters of galaxies, and about 5\% other objects.  The error circle 
	associated with the position of an X-ray source depends on the instrument used, the 
	characteristics of the field, and the strength of the source. Positional information available 
	from central regions of the field of view of imaging X-ray telescopes (e.g., ROSAT) 
	usually has errors ranging from about 5 arcsecs to 20 arcsecs depending on the type of detector 
	used at the focus of the telescope. With error circles of 20 arcsecs, there is usually no more 
	than one object brighter than 19.5 mag (Pandey et al. 2005) in the V-band within the error circle, 
	thus targets can be identified very easily using images taken with a CCD on a 3.6 m telescope and 
	taking short exposures of 10s of seconds to a minute. Many new soft X−ray sources have recently been 
	discovered in surveys with the Einstein and ROSAT. A large fraction of these objects are probable new 
	accretion  binaries, either magnetic CVs or Low Mass X-ray Binaries. Our aim will be to identify 
	the optical counter parts and then to carry out detailed optical/near-IR photometric and 
	spectroscopic studies. The optical follow-up observations of the ASTROSAT sources would also be 
	a key target for the 4K$\times$4K CCD Imager (Sreekumar 2005).

	\subsection{Optical identifications of radio sources from GMRT:}
	Giant Metrewave Radio Telescope (GMRT) has the potential of placing the country in a dominant position 
	in the world of low-frequency radio astronomy. Optical observations of some of these radio sources
	with the 3.6m DOT is another obvious scientific goals (Samuel \& Ian 2011). 
	Finding the optical/near-infrared counterparts of these radio sources is going to be the first essential 
	step in ascertaining their nature and isolating the most promising objects among them, which 
	could then be taken to 6-10m class telescopes for a detailed study. Without these optical 
	identifications, the science targets of GMRT would remain largely unfulfilled and thus a 
	guaranteed access to medium to large-size optical telescopes are required. 
	The CCD Imager along with the modern 3.6m DOT could provide 
	sufficiently deep detections (B$\sim$ 25 mag) within an hour of exposure, thus yielding optical 
	identifications for a majority of the radio sources detected with GMRT (Swarup et al. 1991
	Singal et al. 2004, Cameron et al. 2005).

	\section{4K$\times$4K CCD Imager}\label{sec:intro}

	The 4K$\times$4K CCD Imager with motorized filter-housing and camera mounting arrangements
	is designed to be mounted at the axial port of the 3.6m DOT. The f/9 beam of the
	telescope system is directly used without any focal reducer and has a plate-scale
	of $\sim$ 6.4 arcsec/mm. It is planned to use the f/9 beam directly to utilize the central
	unvignated $\sim$ 10$\times$10 arcmin$^2$ of the science field using appropriate filters.

	\noindent
	The design of the Imager and its various sub-components were prepared indigenously at ARIES.
	A suitable blue-enhanced 4K$\times$4K CCD camera with a pixel size of 15 micron and associated
	ARCHON control electronics was purchased in 2012 as demanded for the proposed 
	science goals and availability of such CCDs in the market.
	The required mechanical design, manufacturing and filter wheel motorization 
	were completed by a team of engineers and scientists at ARIES. A schematic 
	block diagram of the CCD Imager is shown in Fig. 1. The details about development of various 
	phases of the 4K$\times$4K CCD Imager project are described below.

	\begin{figure}[h]
	\begin{minipage}{8cm}
	\centering
	\includegraphics[width=8cm]{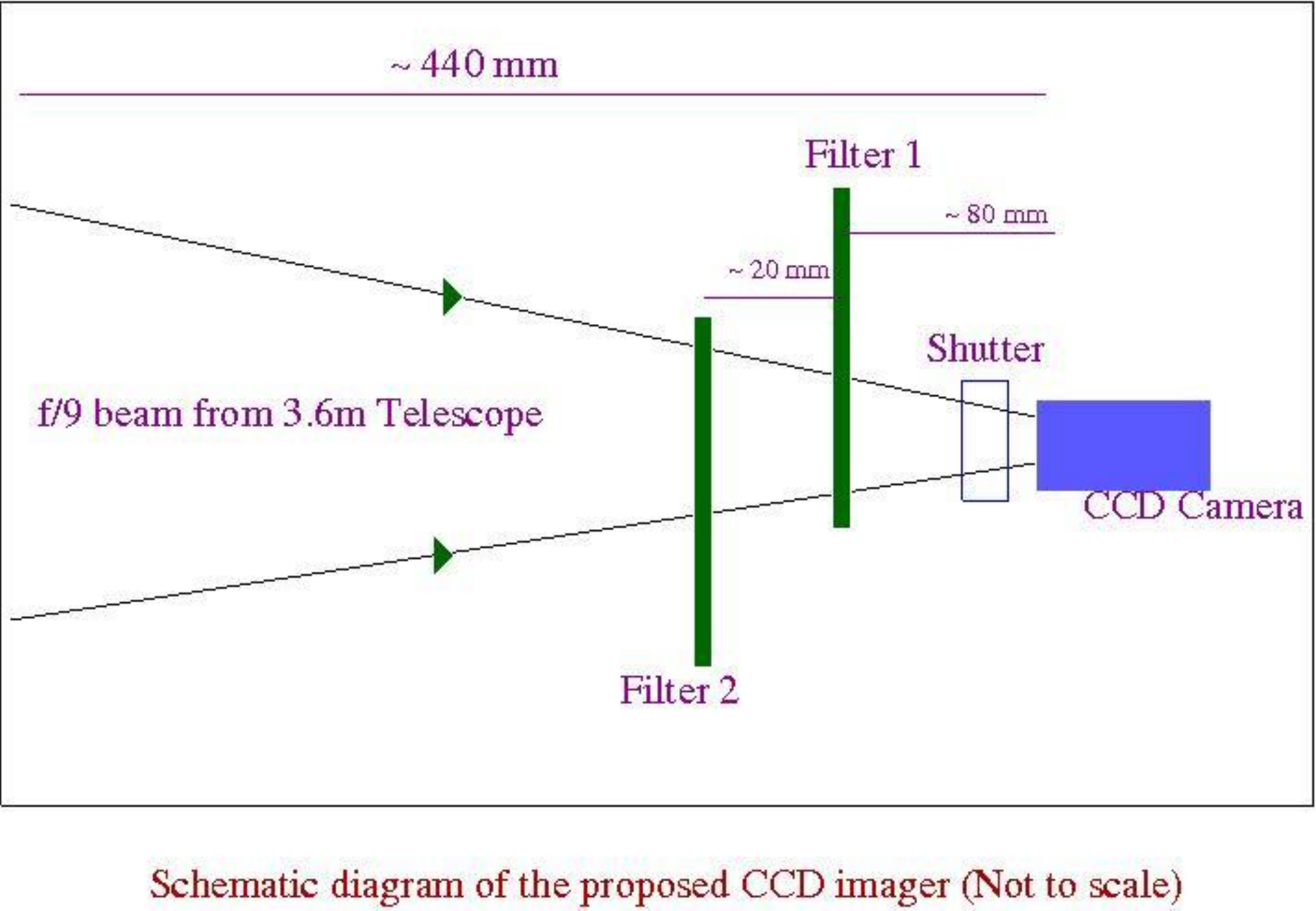}
	\caption{Schematic diagram of the 4K$\times$4K CCD Imager.} 
	\end{minipage}
	\hfill
	\begin{minipage}{8cm}
	\centering
	\includegraphics[width=8cm]{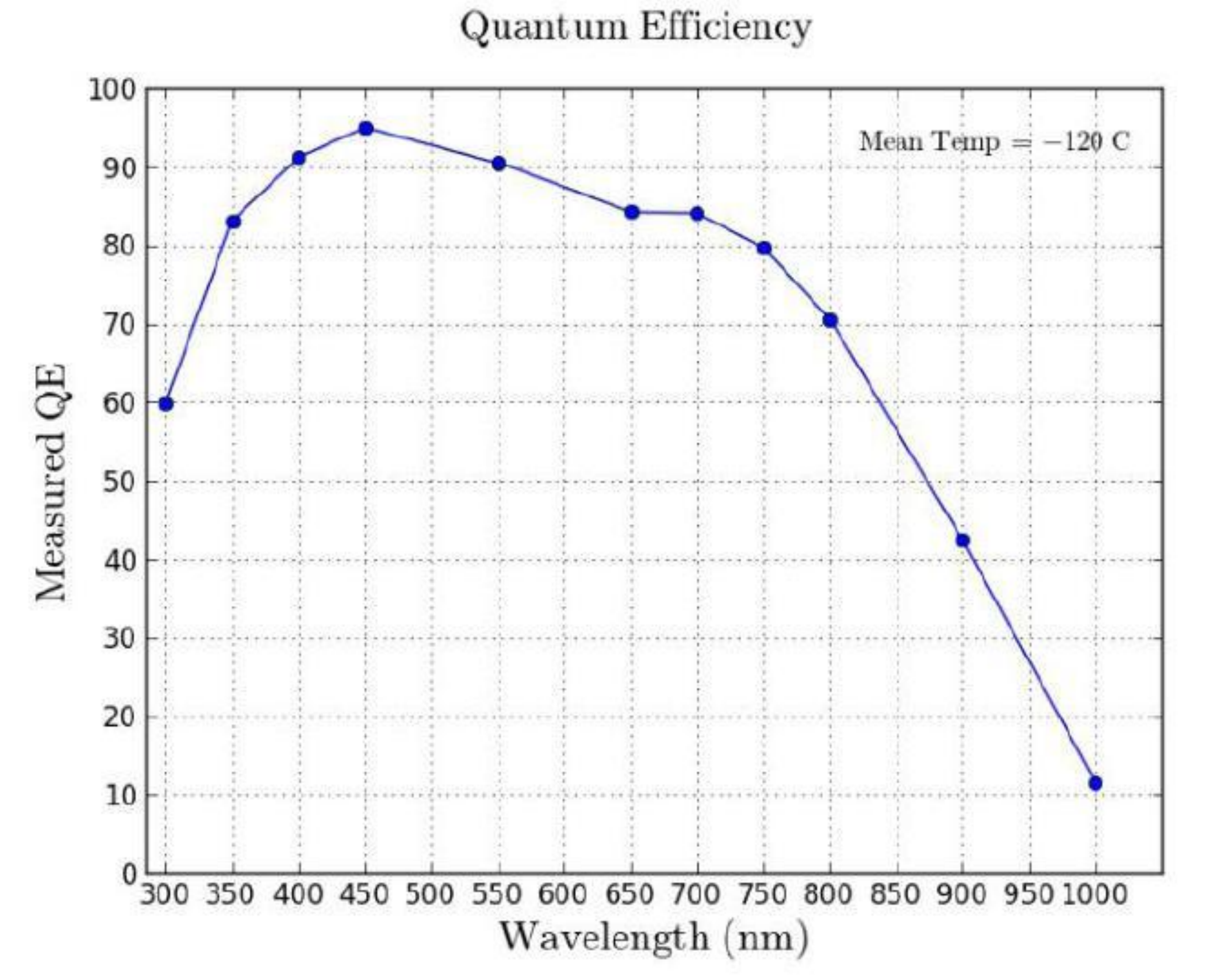}
	\caption{Measured values of the quantum efficiencies as a function of wavelength for the STA 4K$\times$4K CCD chip.} 
	\end{minipage}
	\end{figure}

	\subsection{Zemax Analysis}\label{sec:intro}

	The f/9 beam at the axial port of the 3.6m DOT was used to analyze the image
	quality due to insertion of filters, possible tilt of the filters and
	tilt of CCD chip plane using Zemax software\footnote{http://http://www.zemax.com//}. 
	One of the two filters (fused silica with
	5 mm thickness) is inserted in the optical path of the telescope before the
	focal plane. Focal plane of the telescope is shifted by 1.58 mm away from the
	actual focal plane due to 5 mm filter thickness whereas change in the image quality
	is insignificant. This image shift can be compensated by moving secondary
	mirror in z-axis direction. Filters were also tilted by 1 deg individually and image
	quality is observed. No significant change in the RMS radii is observed.
	There is a change of around 4 microns in RMS radii is observed for the tilt of
	5 deg. It is to be note that the tilting is done in one direction only by assuming that the
	system is symmetric.

	\noindent
	Further, CCD plane was tilted by around 11 arc-min (equivalent amount of flexure) under 
	the Zemax environment and image quality was observed. As a result, a change of 4-5 microns 
	in the RMS radii at the extreme/corner field points was noticed which goes significant if the
	tilt is more. The spot diagrams related to the present analysis are described below in 
	Figs. 3 and 4.  

	\begin{figure}[h]
	\begin{minipage}{8cm}
	\centering
	\includegraphics[width=8cm]{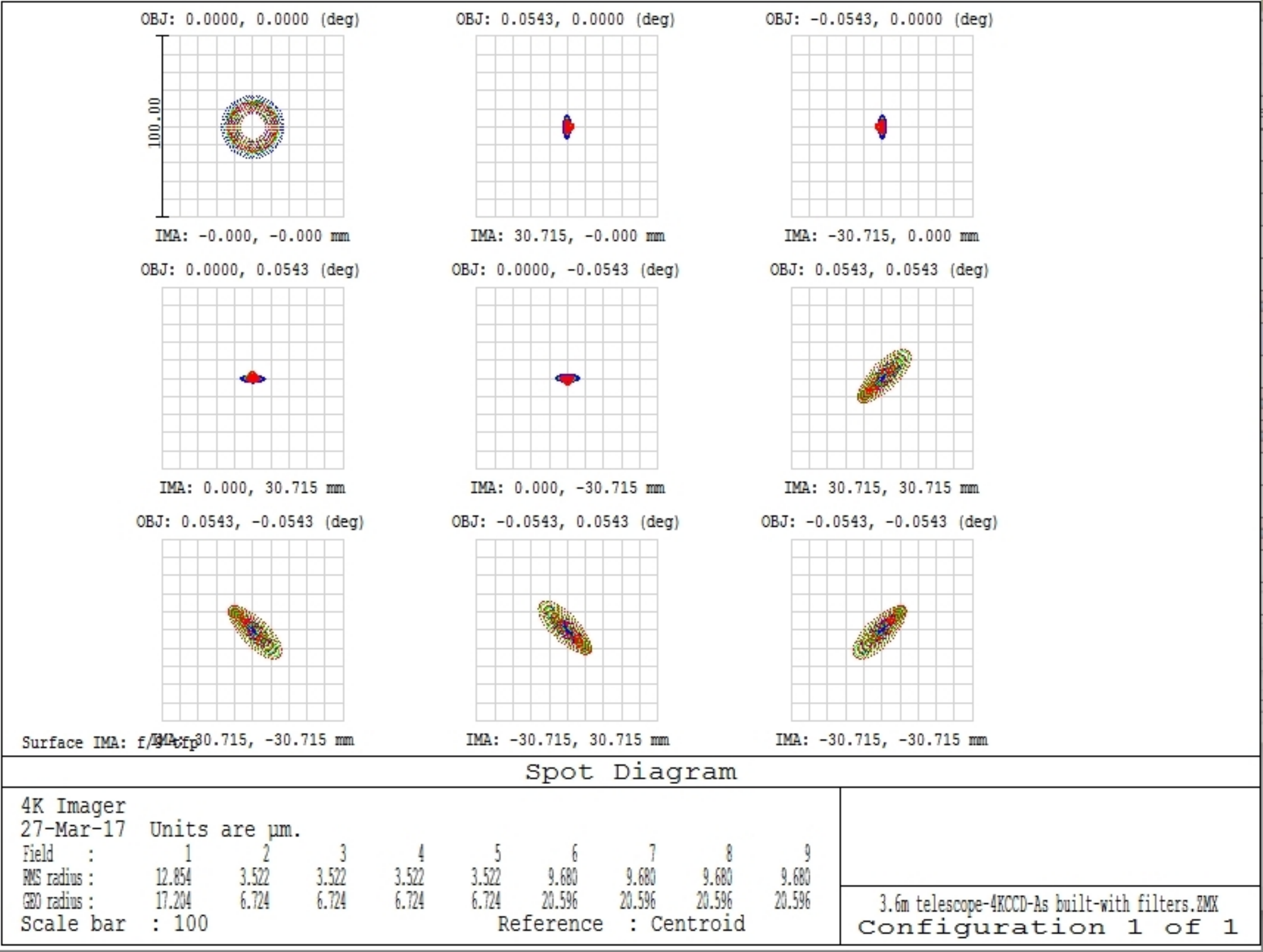}
	\caption{Spot diagram of the 3.6m DOT as produced by the Zemax software using the
	f/9 beam at the axial port with 5 mm thick filter placed in the beam path
	with focus compensated by secondary mirror. \label{fig_2}}
	\end{minipage}
	\hfill
	\begin{minipage}{8cm}
	\centering
	\includegraphics[width=8cm]{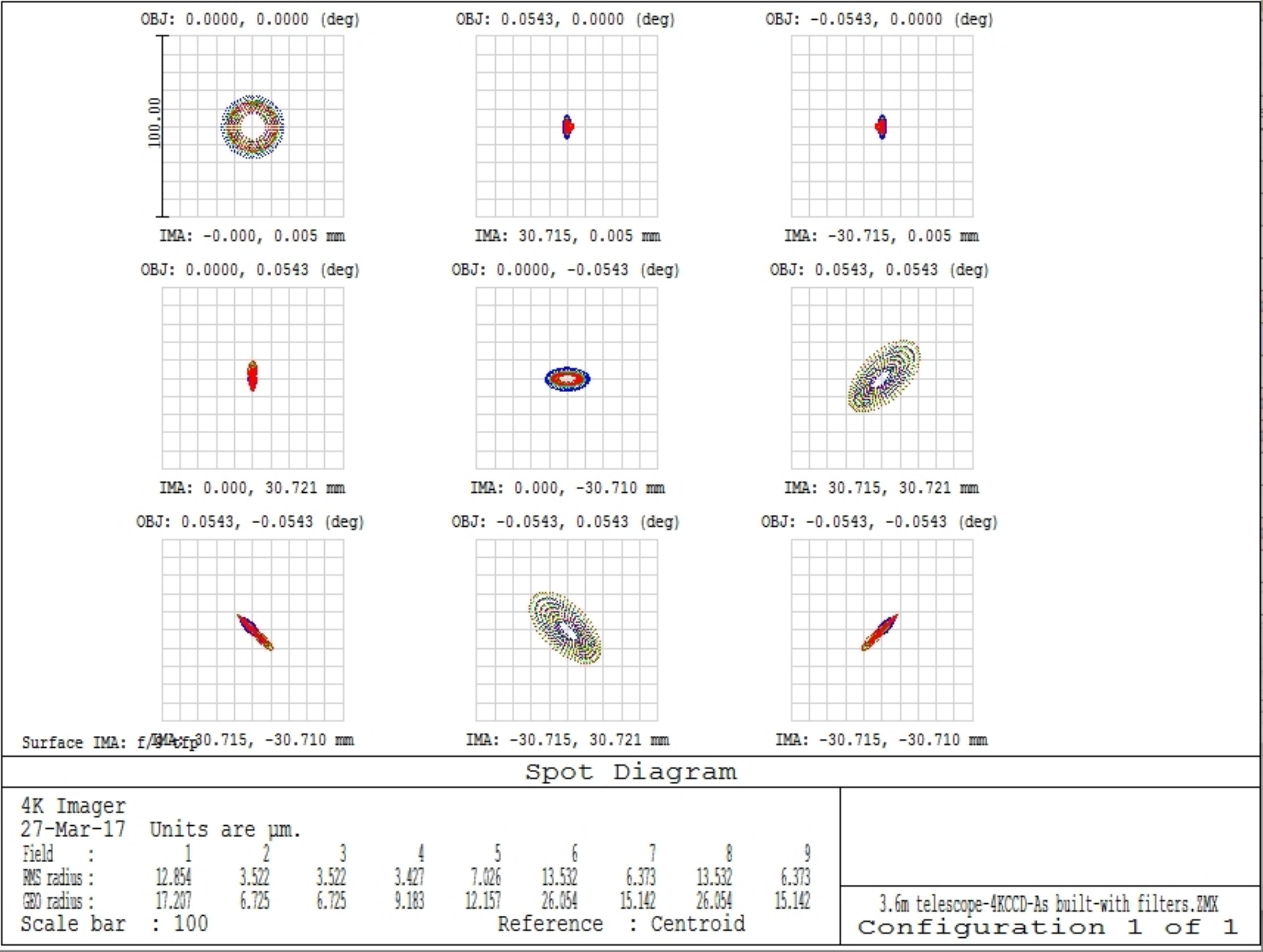}
	\caption{Spot diagram of the 3.6m DOT as produced by the Zemax software using the
	f/9 beam at the axial port with 5 mm thick filter inserted in the path
	but the CCD plane tilted by 11 arcmin with respect to the normal to optics
	axis.\label{fig_3}}
	\end{minipage}
	\end{figure}

	\subsection{The CCD detector}\label{sec:intro}

	The requirements of a CCD detector/pixel sizes are mainly governed by the telescope optics and the
	seeing conditions of the site. At the time of purchase, pixel size of the large format astronomical
	grade blue-enhanced single chip CCDs in the market were limited to 15 micron. At Devasthal, 
	the median seeing is $\sim$ 1.1 to 1.2 arcsec and it was proposed to use the f/9 beam directly at the 
	axial port of the 3.6m DOT. So, based on the scientific goals mentioned above, it was decided to 
	purchase a blue-enhanced, back-illuminated 4K$\times$4K CCD chip with a pixel size of 15 micron in 2011
	from Semiconductor Technology Associates Inc. (STA) USA. The technical parameters of the STA CCD 
	chip/camera are listed in Table 1 (for more details about the camera please refer website of 
	STA\footnote{http://www.sta-inc.net/}). For the STA CCD chip, the quantum efficiency curve as a function 
	of wavelength is also shown in Fig. 2. 

	\begin{table}
	\caption{Characteristic properties of the 4K$\times$4K CCD chip STA 4150 BI (number 155727 W12 D01) 
	are summarized in this table.}
	\vspace{0.5cm}
	\small
	\begin{center} 
	\begin{tabular}{| c | c |}
	\hline 
	Specifications/&Values \\
	Parameters& \\
	\hline
	CCD Chip & 15 micron/pixel, 4096x4096 pixels, back-illuminated, 16-bit A-D\\
	Full-well capacity & 250k -- 265k electrons, MPP/non-MPP/CAB modes  \\
	Gain & 1,2,3,5,10 electrons/ADU (selectable) \\
	Read-out Noise (@Speed) & 7-9 electrons (@1 MHz) or 4-6 electrons (@500 KHz) or
	2-3 electrons (@100 KHz) \\
	Linearity/CTE & 10\% to 90\% of the full well / 0.999999  \\
	Binning & 2x2, 3x3 or 4x4 (select-able)  \\
	Dark Current & 0.0005 electrons/sec at -120 C \\
	Dewar/cooling & Liquid Nitrogen cooled, -121.3 C  \\
	\hline 
	\end{tabular} 
	\end{center} 
	\end{table} 
	\noindent
	For the given plate scale of 6.4 arc sec/mm, a 15 micron 4K$\times$4K CCD camera is able
	to image $\sim$  6.5$\times$6.5 arc min of the sky, though images would be over-sampled. 
	So, there is a provision to bin the pixels in
	various combinations to optimize with seeing values during observations. Binning
	the pixels further improves the signal to noise and the read out would be
	faster for a given set of read-out noise and gain values. Other sub-components 
	of the 4K$\times$4K CCD camera are described below.

	\subsubsection{The ARCHON controller and communications}

	For the STA CCD, the controller is designed and developed in a customized manner and known
	as ARCHON controller \footnote{http://www.sta-inc.net/archon/}.
	The camera controller software is written in QT Version-5. The ARCHON Controller is
	connected to Port 1 \& 2 of LIU-2  by a LC-LC multi-mode patch cord. LIU-2 is
	connected to LIU-1 through a 12 core OFC which runs through azimuth cable wrap.
	 HP Gigabit switch is connected to LIU-1 by a LC-LC patch cord and finally the
	Archon Imager Software PC is connected to HP Gigabit switch by a cat 6
	Ethernet cable through its 1 GB per/sec Ethernet port (eno2). An Intel 82572 SFP
	Gigabit Fiber Optical card is also installed in the PC which can also be used
	to connect the controller.  Archon controller is compatible only with Gigabit
	ports and thus it should be kept in mind that Gigabit Ethernet or Fiber port
	should be there in the Software PC. The controller, mounted along with the 
	camera, operates at 220V and 50Hz power supply. The power dissipation of the 
	ARCHON camera controller is $<$ 50 W to avoid any thermal effects near the 
	telescope during observations.


	\subsubsection{Cooling mechanism}
	Liquid nitrogen (LN2) cooling mechanism is used to achieve CCD temperatures as low as $\sim$ -120C to
	minimize the dark current to possible low values, good for longer exposures detecting fainter
	objects. Also, the liquid Nitrogen Dewar along with the camera controller was well-suited to the
	given specifications of the mounting arrangements. The cooling hold time of the camera dewar is
	kept of the order of $\sim$ 15 hours to get a stable temperature during observations. Before filling 
	LN2, the dewar is evacuated using Edwards turbo molecular pump below 5 mTorr level. 
	Generally, the dewar takes around 2 hours to reach -120 C, the suggested temperature during observations. 
	Separate manuals are provided for further assistance about LN2 filling and possible precautions.

	\subsubsection{Shutter}
	The 4K$\times$4K CCD Imager system also needs a shutter mechanism to control the light arriving
	at the science detector in a given exposure time. One important characteristics of the shutter
	is its ability to operate uniformly, so that every part of field obtains the same length of
	exposure particularly in case of shorter exposures. In case of the 4K$\times$4K CCD Imager, it was
	decided to use 125 mm Bonn shutters with a mechanism to run the drivers of the shutter assembly
	through the CCD controller software\footnote{http://www.bonn-shutter.de/shutter\_types.php}.

	\subsection{The mechanical assembly of the Imager}
	The prime requirement of the opto-mechanical design of the Imager assembly is in particular to
	mount the optical filters, the CCD camera to hold and move these elements with sufficient
	precision while the telescope is in motion. The opto-mechanical design of the Imager assembly was
	simplified as much as possible due to the space limitations, manufacturing complexities and
	given weight constraints. The filter wheel assembly of the Imager was mounted at the axial port of 
	the 3.6m DOT with three spider arms connected to the 2-ton dummy weight
	structure through a baffle arrangement to prevent stray light. The round discs of the two filter
	wheels with six slots to mount the set of filters are positioned along with the motors, gears and
	couplers for an easy accessibility for  maintenance, cleaning and changing of filters
	periodically. In the design of the optical mountings, it is important that the clamping forces
	that constrains the position of the optical filters and it should not be too large. Mounting
	stress on the optics can cause surface deformations or birefringence in the optical material,
	could degrade the image quality performance. To avoid these possible issues all the 
	filters are supported by POM (polyoxymethylene) material. While preparing the mechanical design, 
	care was taken to minimize the effects like differential
	thermal expansion of the optical and mechanical parts to reduce the possible mounting stress.
	Finite-element analyses (FEA) were performed to verify that stiffness deformation of the filter
	wheel assembly due to gravity or the mounting stress on optics. A free version of the analysis
	software package called ``Ansys'' was used for the FEA analysis and related calculations.

	\noindent
	The mounting arrangements for this instrument were designed considering the available space
	at axial port of the cassegrain focus of the 3.6m DOT. Mechanical specifications of the 3.6m DOT 
	dictates that the maximum allowable
	instrument weight at the cassegrain axial port must be below 2000 kg  with a center of gravity
	located 800 mm below the instrument flange. Figure 5 shows the full mechanical 
	structure of the of the 4K$\times$4K CCD Imager mounted at the instrument 
	flange. More details about various sub-components of the Imager are given 
	in the following subsection.

	\subsubsection{Angle Brackets}
	Angle brackets connect the housing structure to telescope axial dummy weight. The three brackets are 
	very stiff components placed at an angle of 120 degrees and over all load distribution and stiffness 
	of instrument depends on the three angle brackets. These angle brackets are mounted at 120 degree each,
	connected with the dummy weight and the filter wheel housing at both ends 
	and are made of aluminum T6 6061 alloy.

	\subsubsection{Filter Housing}
	The requirement of Filter housing structure is to house and support the components like filter wheel
	assembly, drive mechanisms, CCD camera and control electronics etc. The filter housing structure
	should be rigid and stiff enough to take overall weight and other forces acting on it while
	observations and in idle conditions. Apart from supporting the sub-components, the housing structure
	has a provision to access the filters for maintenance and cleaning of filters periodically.
	Two identical hollow bearing shaft of different lengths help to maintain gap between the filter wheels 
	and supporting the filter wheels. The bearing shaft is made of hardened steels and ground to tolerance 
	value of j6 for bearing fittings. The size of bearing shafts are 70 mm  and weights of $\sim$ 1 kg each.
	Specialized deep groove ball bearing are used for filter wheel drive for a smooth and jerk free rotation
	of the two filter wheels. External and internal diameters of deep groove ball bearings are 100 mm and 
	70 mm respectively with a breadth of 16 mm. 

	\noindent
	The housing is combination of machined plates made up of aluminum T6 6061 alloy material. Full
	size of housing is approximately 833$\times$830$\times$117 mm$^3$. The sole purpose of the design and
	building plate box structure is to ease the alignment related issues and to address maintenance
	related requirements. The housing has unique plates at diagonal locations to support and hold
	the stepper driving mechanisms. Top and bottom plates are used to simply support filter assembly and
	the CCD camera assembly.

	\subsubsection{Filter wheel assembly}
	Prime requirement of filter wheel is to accommodate 10 optical filters and rotate 
	position filters as per the control or desired by observers at a time of 
	observation. Filter wheel is one of the high precise components of this instrument due to requirement 
	of accurate positioning and supporting of filters with low mounting stress. To achieve these 
	requirements, the filter wheels are built by reducing weights 
	as much as possible and precisely cutting the rectangular pockets, slots
	to accommodate filters. The outer periphery of the filter wheels are cut by module-1 teeth in
	precise gear hobbing machine and bearing hole machined at center. To avoid inertial play and stiffness
	problems materials were scooped while manufacturing the filter wheels. Each filter wheel has 6 slots of 
	size 90$\times$90 mm with a clear aperture of 85$\times$85
	mm. One slot in each filter wheel will be empty to use only one filter at the time of observation.
	The diameter of each filter wheel assembly is 434 mm mounted along with SKF bearings and bearing covers 
	for mounting purposes.

	\begin{figure}[h]
	\begin{minipage}{8cm}
	\centering
	\includegraphics[width=8cm]{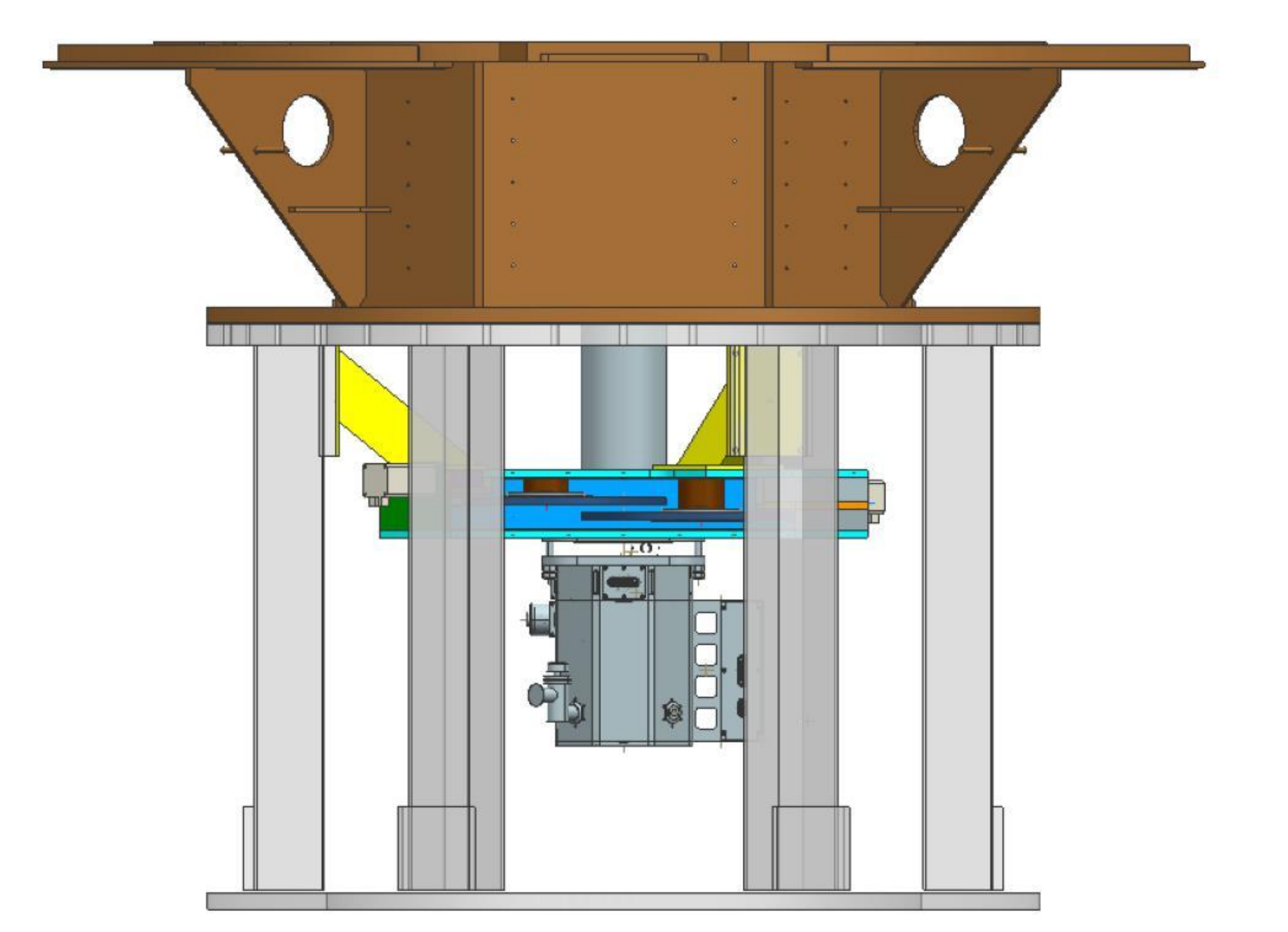}
	\caption{Schematic diagram of the proposed CCD Imager within the given dummy
	weight and the instrument envelop. The full assembly including angle brackets, 
	the filter wheel case, motors/gears and the CCD camera 4K$\times$4K CCD camera
	are shown in the figure as produced by the software called ``pro-e''.}
	\end{minipage}
	\hfill
	\begin{minipage}{8cm}
	\centering
	\includegraphics[width=8cm]{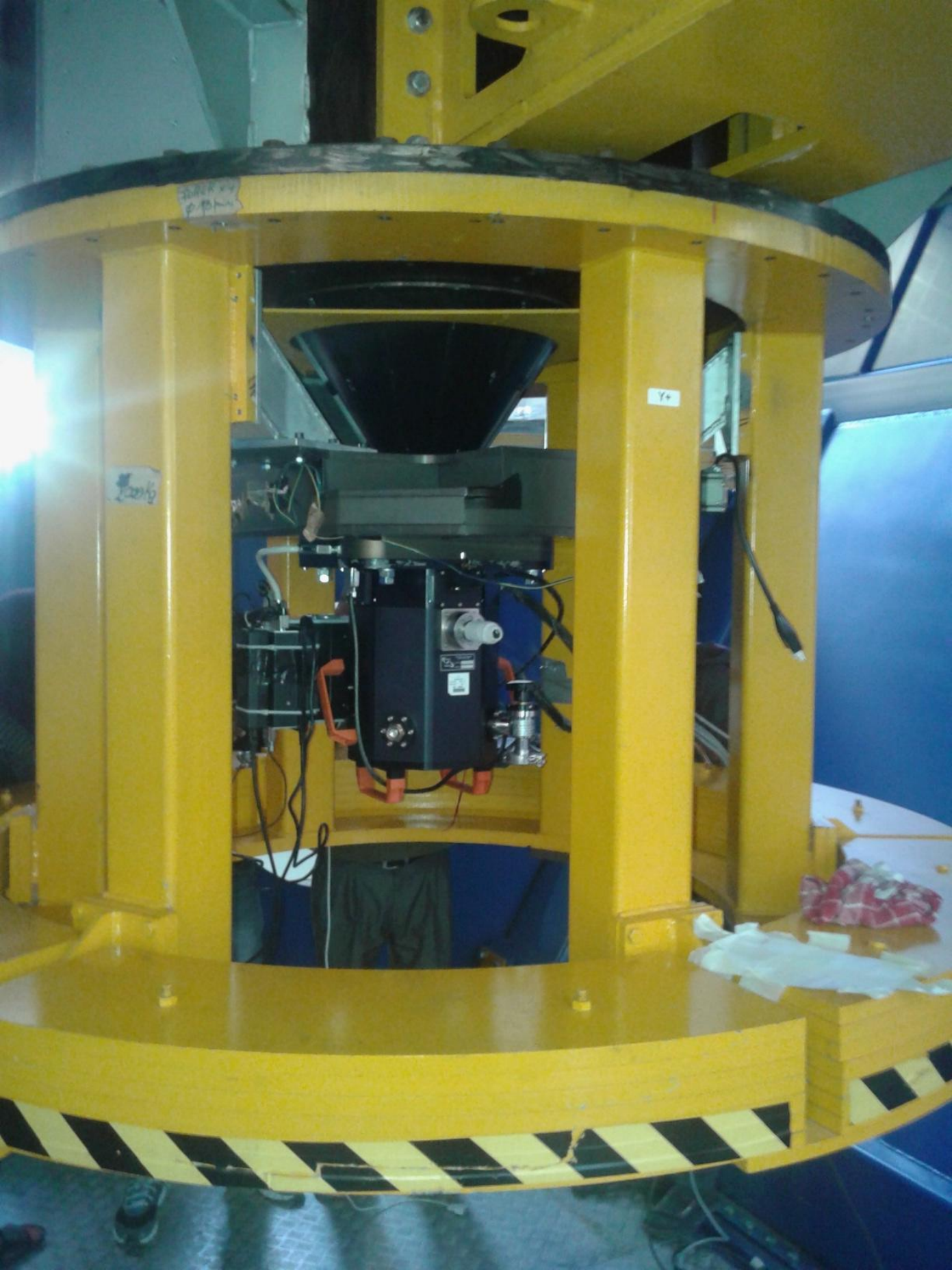}
	\caption{The fully assembled 4K$\times$4K CCD camera along with shutter
	and automated filter wheels mounted at the axial port of the 3.6m DOT
	on 9th December 2015.} 
	\end{minipage}
	\end{figure}

	\subsubsection{ Filter wheel Drive system}
	Each filter wheel accommodates five filters at a time and each filter is 
	located at 60 deg index angle. As required during observations, each filter 
	is rotated and positioned several times by maintaining precise tolerances
	limits. The rotation of filter wheels is controlled by control electronics 
	through pinion gear, worm gear box (ratio 30, anti-backless Moffet), 
	electronics sensors attached between the filter wheel and stepper motor drive. 
	The total gear reduction ratio between motor to wheel drive is 540, reducing 
	error between motor to filters approximately by a factor of 540. The required 
	repeatability accuracy of each filter wheel (a few tens of microns) is 
	achieved through anti-backlash gear fitted between filter 
	wheel and the gear box. Muffet gear box, used in the Imager, is a 
	less weight compact precise anti-backlash worm gear system manufactured by 
	muffet inc. UK. The pinion gear used is module-1 teeth spur gear made of high 
	quality nylon material. The purpose of using nylon material is avoiding 
	residual stress and wear between aluminum filter wheels teethes and pinion 
	teethes. The number of teethes in pinion is 24. The stepper motor was used 
	to rotate filter wheels with the desired precision. The motion of the motors 
	were controlled through micro-controller based card and were interfaced 
	with a GUI based s/w provided to control the CCD electronics.

	\subsubsection{Filters}
	There is a provision to use both SDSS $ugriz$ and Bessel $UBVRI$ broad-band 
	filters (in the spectral range 360-1000 nm) for imaging purposes along with 
	the 4K$\times$4K CCD Imager. As per the design of the Imager and with the 
	given space at the mounting flange, the f/9 beam of the telescope required 
	a filter sizes of $\sim$ 90 mm each. Central wavelengths and band-width of 
	each of the 10 filters in use are tabulated in columns 1 and 2 of the table 2. 
	The transmission 
	coefficients of the all 10 filters were verified in the lab and were found as 
	per specifications as tabulated in column 3 of the table 2. 

	\section{Motorized Filter wheel}\label{sec:intro}
	Implementation of motorized filter wheels is one of the major in-house 
	activity of the CCD Imager project. The mechanisms (both software and hardware) were 
	designed, developed and implemented solely by ARIES staff members involved 
	towards this project. A brief description about this automation is
	outlined below.

	\begin{table}
	\small
	\caption{In this table, specifications of all 10 broad-band filters including 
	central wavelengths, band-width, transmissions are tabulated. The measured 
	values of the atmospheric extinction coefficients in all the 10 filters using the 
	combination of the CCD Imager and the 3.6m DOT are 
	tabulated in the last column.}
	\begin{center} 
	\begin{tabular}{| l | l | l | l |}
	\hline 
	Filters&Central wavelength/& \% Transmission&Extinction coefficients \\
	       &Band-width (\AA)         &(measured values)& (measured values in mag) \\
	\hline
	Bessel U & 3663 / 650& 62.0 & 0.64$\pm$0.03\\
	Bessel B & 4361 / 890& 70.0 & 0.39$\pm$0.02 \\
	Bessel V & 5448 / 840& 80.0 & 0.29$\pm$0.02 \\
	Bessel R & 6407 / 1580&88.0 & 0.22$\pm$0.01 \\
	Bessel I & 7980 / 1540&80.0 & 0.17$\pm$0.02 \\
	SDSS $u$ & 3596 / 570 &63.0 & 0.70$\pm$0.02\\
	SDSS $g$ & 4639 / 1280&88.0 & 0.30$\pm$0.04 \\
	SDSS $r$ & 6122 / 1150&85.0 & 0.20$\pm$0.02 \\
	SDSS $i$ & 7439 / 1230&87.0 & 0.14$\pm$0.02\\
	SDSS $z$ & 8896 / 1070&88.0 & 0.09$\pm$0.03 \\
	\hline 
	\end{tabular} 
	\end{center} 
	\end{table} 

	\noindent
	3.6m DOT Imager instrument consists of two filter wheels. Both filter wheels have separate set of 
	six filter positions namely $U, B, V, R, I, C$ (Clear) and $u, g, r, i, z, c$ (Clear), respectively.
	Microcontroller based control unit and GUI software are used for the positioning of two filter wheels
	in the Imager Instrument. Control unit consists of a PIC microcontroller having Serial Communications
	Interface (SCI) module USART (Universal Synchronous Asynchronous Receiver Transmitter) and a circuit
	that converts from RS-232 compatible signal levels to the USART's logic levels and vice-versa. The 
	control unit communicate with the interfacing PC
	over TCP/IP protocol using simple ASCII commands through RS-232 to Ethernet converter and vice versa.
	For moving the wheels stepper motors along with suitable drivers/amplifiers have been employed.
	Homing is achieved after powering ON using Hall Effect sensors. A detailed log of commands, status
	and errors are continuously generated by the GUI software. Both the control unit and the software 
	have been successfully tested and integrated with the Imager instrument.

	\noindent
	Initially at the POWER ON, homing is achieved automatically and both filter wheels are positioned at
	their clear position. Whenever any command is received by the controller,
	request of command is executed first. Controller will be ready for the execution of next command
	only after the execution of previous command. Any request received from the interface PC will be
	acknowledged after the execution of request by the control unit. Once the GUI of filter wheels is 
	enabled and the CONNECT button is triggered, GUI sends a POWER ON status request to the control unit.
	Communication status between interface PC and control unit is displayed depending on the
	acknowledgment status. Once the communication link is established with the control unit and POWER ON
	status is acknowledged, GUI sends a current POSITION status request to the control unit. After
	acknowledgment from the control unit, current positions of the filter wheels are displayed on the
	screen. Current POSITION status may be requested from interface PC at any time once the control
	unit power is energized. GUI has a provision to request for any individual wheel position movement.
	If any individual position of 1st filter wheel is requested then before moving the 1st filter wheel
	to its requested position, 2nd filter wheel will be moved to its clear position first if it is not
	at its clear position and vice versa. During CCD read out time also filter wheels movement may be 
	requested for next required filter position.

	\begin{figure}[]
	\begin{minipage}{8cm}
	\centering
	\includegraphics[width=\columnwidth]{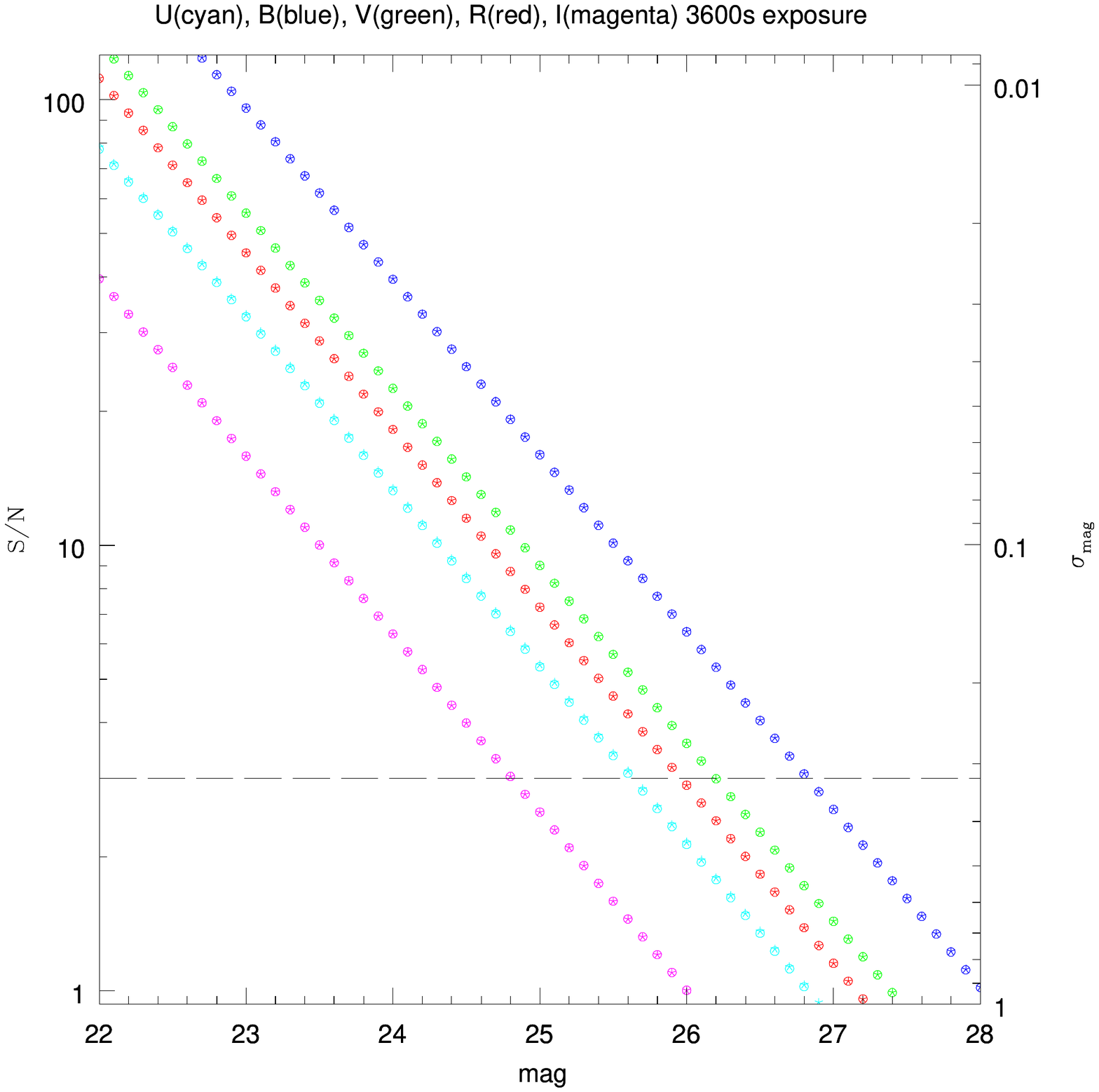}
	\caption{A simulated plot of magnitude (X-axis) Vs. signal-to-noise ratio
	(Y-axis, left) and corresponding error in the magnitude determinations
	(Y-axis, right) based on the throughput calculations (Mayya 1991)
	for the combination of the 3.6m DOT and the 4K$\times$4K CCD camera
	with a pixel size of 15 micron. The plot is reproduced for the set of
	Bessel $UBVRI$ filters and equivalent exposure time of 1 hour each. The
	value of the seeing is assumed to be 1.1 arc-sec along with other
	standard values of parameters given in above table.}
	\end{minipage}
	\hfill
	\begin{minipage}{8cm}
	\centering
	\includegraphics[width=\columnwidth]{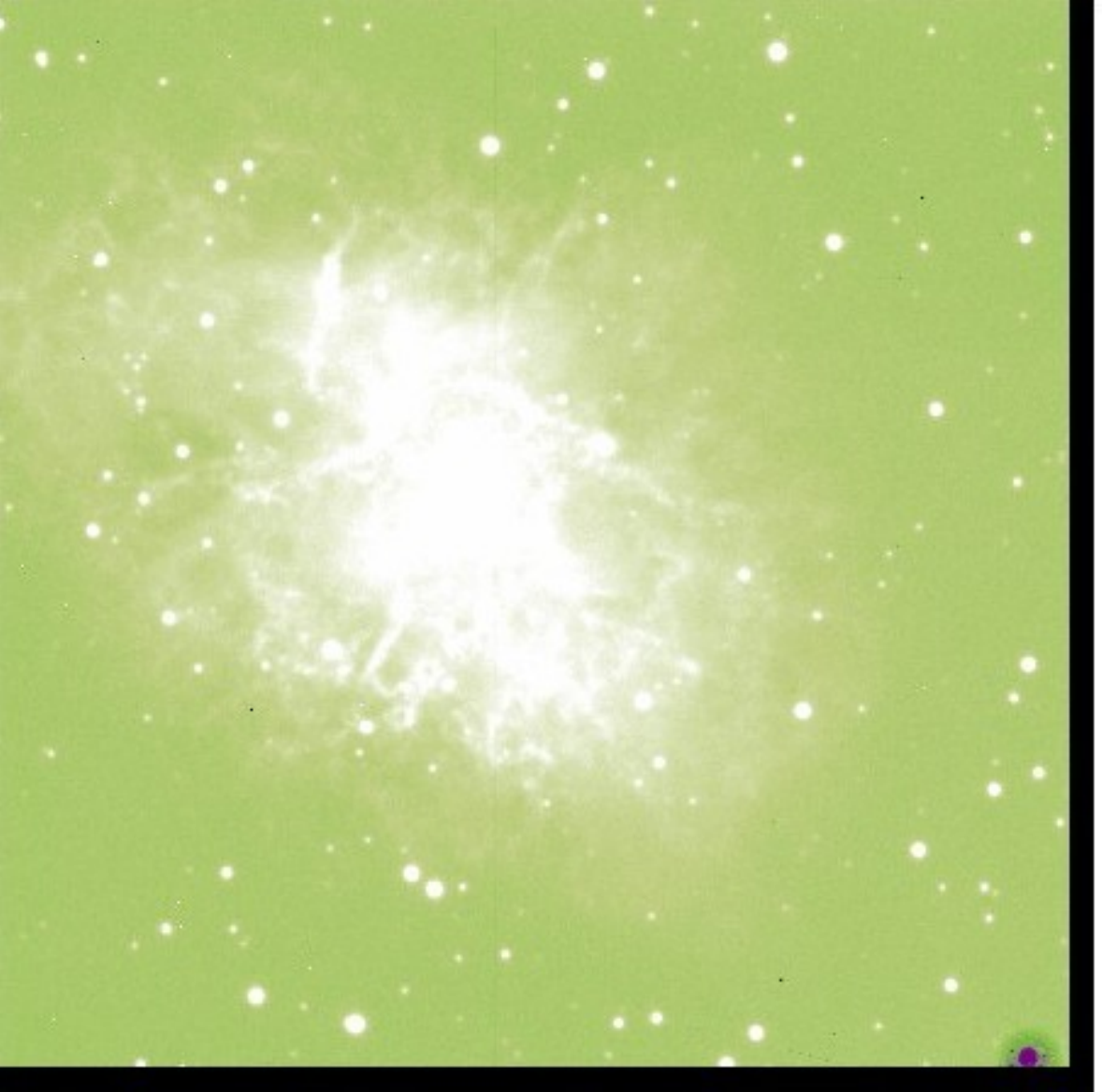}
	\caption{A V-band image of the Crab nebula (a supernova remnant) with an exposure 
	time of 180-sec taken using the 4K$\times$4K CCD camera mounted at the axial port 
	of the 3.6m DOT on 11 December 2015 during testing phase.} 
	\end{minipage}
	\end{figure}

	\section{Throughput Simulations}\label{sec:intro}
	Scientific performance of an instrument depends on maximization of its 
	throughput. The system throughput of the telescope (without a detector like 
	CCD) is expected to be  about 60\% for imaging mode. 

	\noindent
	The expected counts from a point source for the given aperture of the 3.6m DOT were 
	calculated considering the transmission coefficients from the mirrors (M1 reflecting area $\sim$ 72\%, M1 reflectively $\sim$ 92\% (350 – 850nm), M2 reflectively $\sim$ 86\% (400 – 1000nm)); filters ($\sim$ 62\%, 70\%,88\%,80\%and 88\% in Bessel U,B,V,R and I bands respectively) and the  CCD glass (transmission $\sim$ 98\%).  The sky-brightness values for Devasthal were $\sim$ 22.1, 22.4, 21.3, 20.5 and 18.9 mag/arc-sec$^2$ in Bessel U,B,V,R and I bands 
	respectively whereas atmospheric extinction coefficients were $\sim$ 0.49, 0.32, 0.21, 
	0.13 and 0.08 mag in Bessel U,B,V,R and I bands respectively (Kumar et al. 2000). The 
	considered values of the quantum efficiencies for the STA CCD chip were 
	$\sim$ 75\%, 91\%,96\%,81\% and 60\% in Bessel U,B,V,R and I bands respectively. 

	\noindent
	Assuming that each optical photon will be able to produce a corresponding 
	photo-electron and standard values of the CCD parameters like pixel size, 
	dark current, read out noise, gain along with respective values given in 
	above paragraph were used to calculate expected number of counts both from 
	point sources and sky at zenith for an assumed value of seeing of 1.1 arc-sec,
	using the formula given in (Mayya 1991, McLean 1989). The signal-to-noise ratio were 
	also calculated for the stars of different brightness (Deep 2011). 
	In Fig. 7, we have plotted such a result for a specific set of parameters
	(as described in above paragraph) for an assumed value of exposure time of 
	3600 sec for reference.

	\begin{figure}[]
	\begin{minipage}{8cm}
	\centering
	\includegraphics[width=\columnwidth]{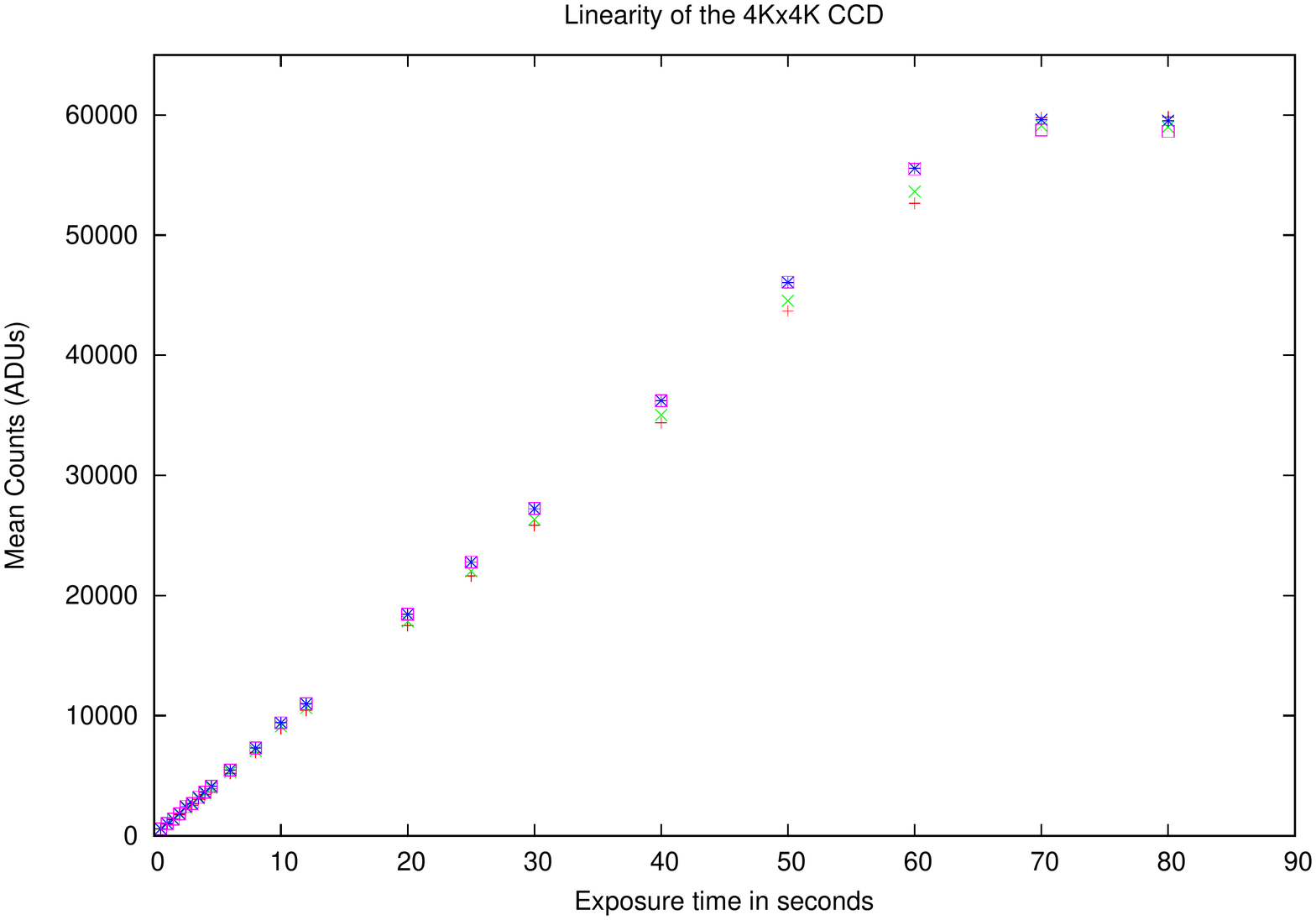}
	  \caption{Linearity plot of the 4K$\times$4K CCD camera showing a linear behavior of the exposure time
	with counts up-to 90\% of the full-well. The four different symbols show the data taken at four different 
	areas within the CCD frame taken at speed of 1 MHz and gain value of 1 electron/ADU.}
	\end{minipage}
	\hfill
	\begin{minipage}{8cm}
	\centering
	\includegraphics[width=\columnwidth]{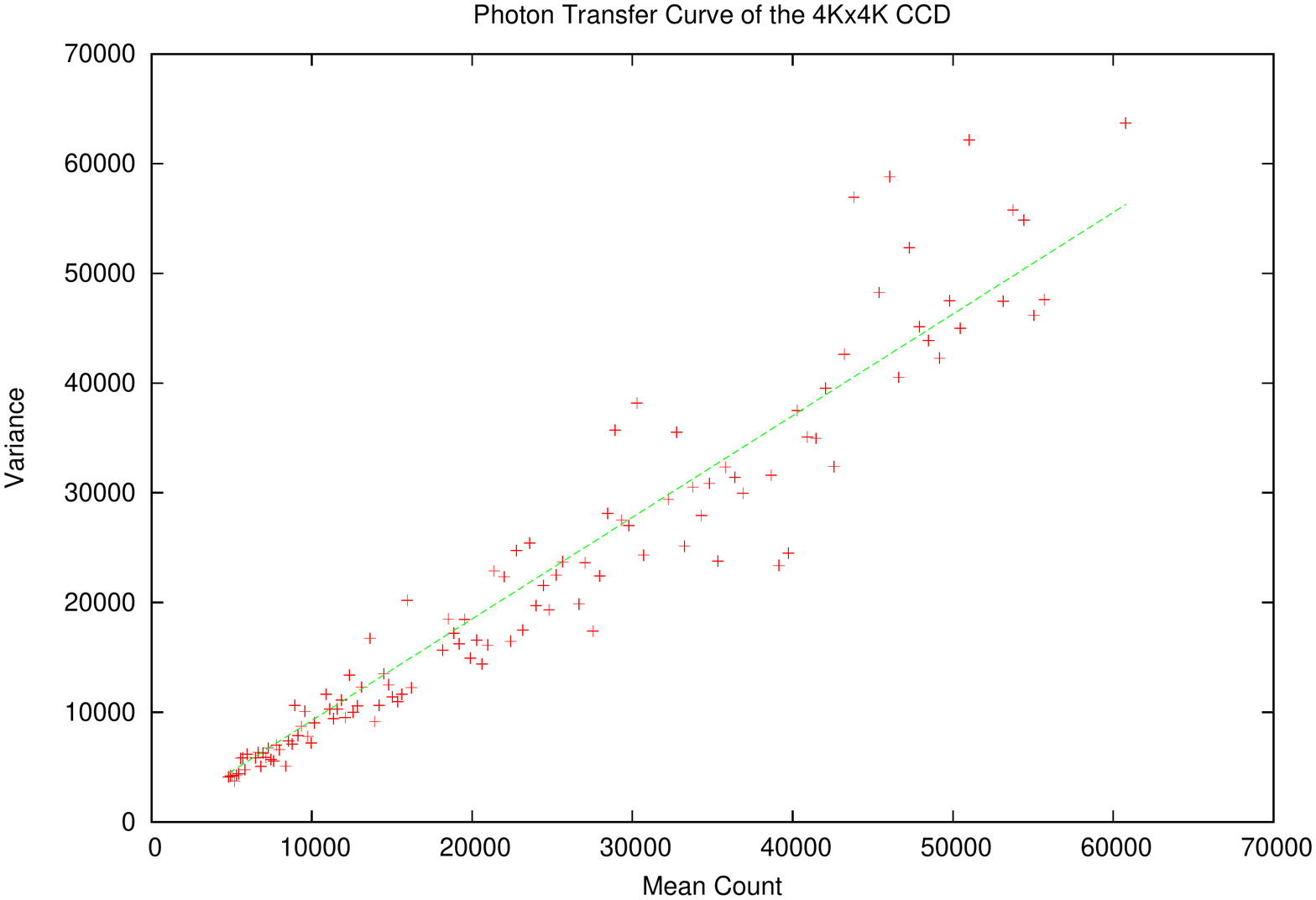}
	 \caption{Photon transfer curve of the 4K$\times$4K CCD Imager taken at gain value of 1 and read out speed of 
	1 MHz. The measured values of gain is 0.93$\pm$0.03 electrons/ADU.}
	\end{minipage}
	\end{figure}

	\section{CCD Characterization}\label{sec:intro}
	The 4K$\times$4K CCD camera was characterized both in laboratory and in open 
	sky conditions while mounted at the axial port of the 3.6m DOT by late 2015.
	Imager mounted at the axial port of the 3.6m DOT on 11 December 2015 is shown
	in Fig. 8. For characterization purposes, several sky flat frames in different filters along
	with bias frames were taken in various combinations of the read-out speeds and 
	gain values select-able through the GUI of the camera control software. In general, 
	over-scan areas were saved along with the frames to get additional 
	information related to the possible fluctuations caused by the electronics.
	Based on the observed data, bias stability was found as per specifications of 
	the LN2 hold time. Standard deviations in observed flat frames were found to be 
	$<$2 \% in general. Linearity of the CCD was verified using the data during November
	2015 and the results are shown in Fig. 9. 

	\noindent 
	Photon transfer curve (PTC) was generated using the flat and bias frames taken at 
	read out speed of 1 MHz and gain setting of 1 electron/ADU as shown in Fig. 10. The 
	slope value between the mean counts and variance (calculated in a small and cleaned
	region of CCD frames) comes out to be 0.93$\pm$0.03, pretty close to the gain value of
	1 electron/ADU. Similar PTCs were also calculated for gain settings of 3 and 5 electron
	/ADU, giving rise to measured values of 2.82$\pm$0.08 and 5.50$\pm$0.25 
	respectively. 

	\noindent
	Gain and read out noise values were also verified occasionally using the two formulas
	as described in Howell (2006). In the formulas below, $\bar{B}$ is the mean value in an area
	of a given bias image, $\bar{F}$ is the mean value in an area of a given flat field,
	$\sigma_{(B_1-B_2)}$ is the standard deviation of pixels into the area of the difference of 
	two bias images and $\sigma_{(F_1-F_2)}$ is the standard deviation of pixels into the area of 
	the difference of the two flat frames.

	\[
	gain= \frac{(\bar{F_1} + \bar{F_2}) - (\bar{B_1} + \bar{B_2})}{\sigma^2_{(F_1-F_2)}-\sigma^2_{(B_1-B_2)}} 
	\]

	\[
	readout \ noise= \frac{gain\times\sigma_{(B_1-B_2)}}{\sqrt{2}} 
	\]

	\noindent
	Using standard routines of IRAF, the derived values of the gain and readout noise values
	using the above two formulas were found to be consistent to those derived using PTCs method.
	The first scientific image of the Crab nebula taken by the 4K$\times$4K CCD
	More details about the characterization of the CCD camera will be
	published separately.

	\begin{figure}[]
	\begin{minipage}{8cm}
	\centering
	\includegraphics[width=\columnwidth]{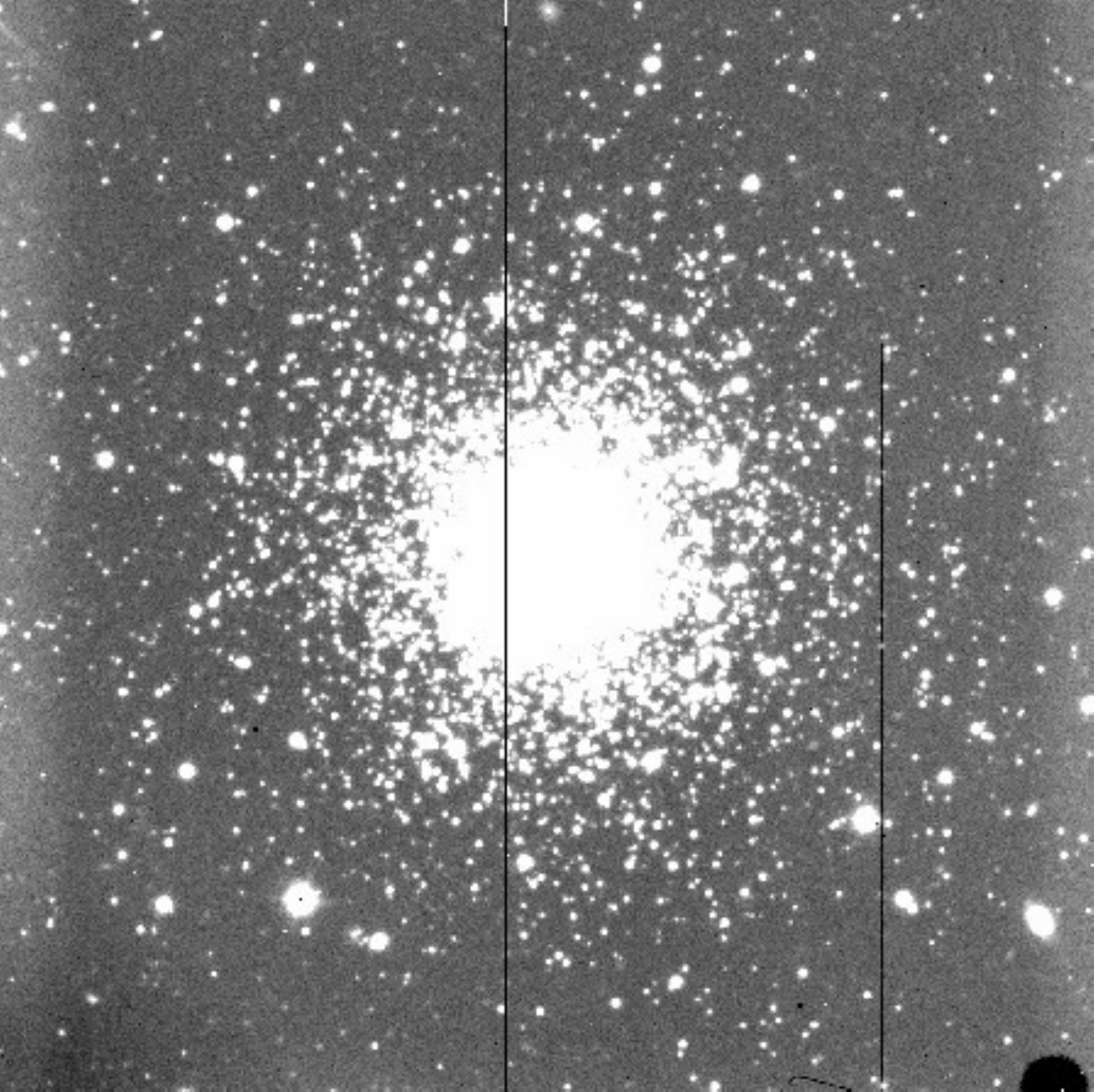}
	  \caption{
	An example of the B-band frame pre-processed (cleaned, trimmed and cosmic-rays removed) of the Globular cluster 
	field NGC 4147 taken in closed loop conditions having a single exposure of 600sec with the 4K$\times$4K 
	CCD Imager mounted at the axial port of the 3.6m DOT on 23rd March 2017.}
	\end{minipage}
	\hfill
	\begin{minipage}{8cm}
	\centering
	\includegraphics[width=\columnwidth]{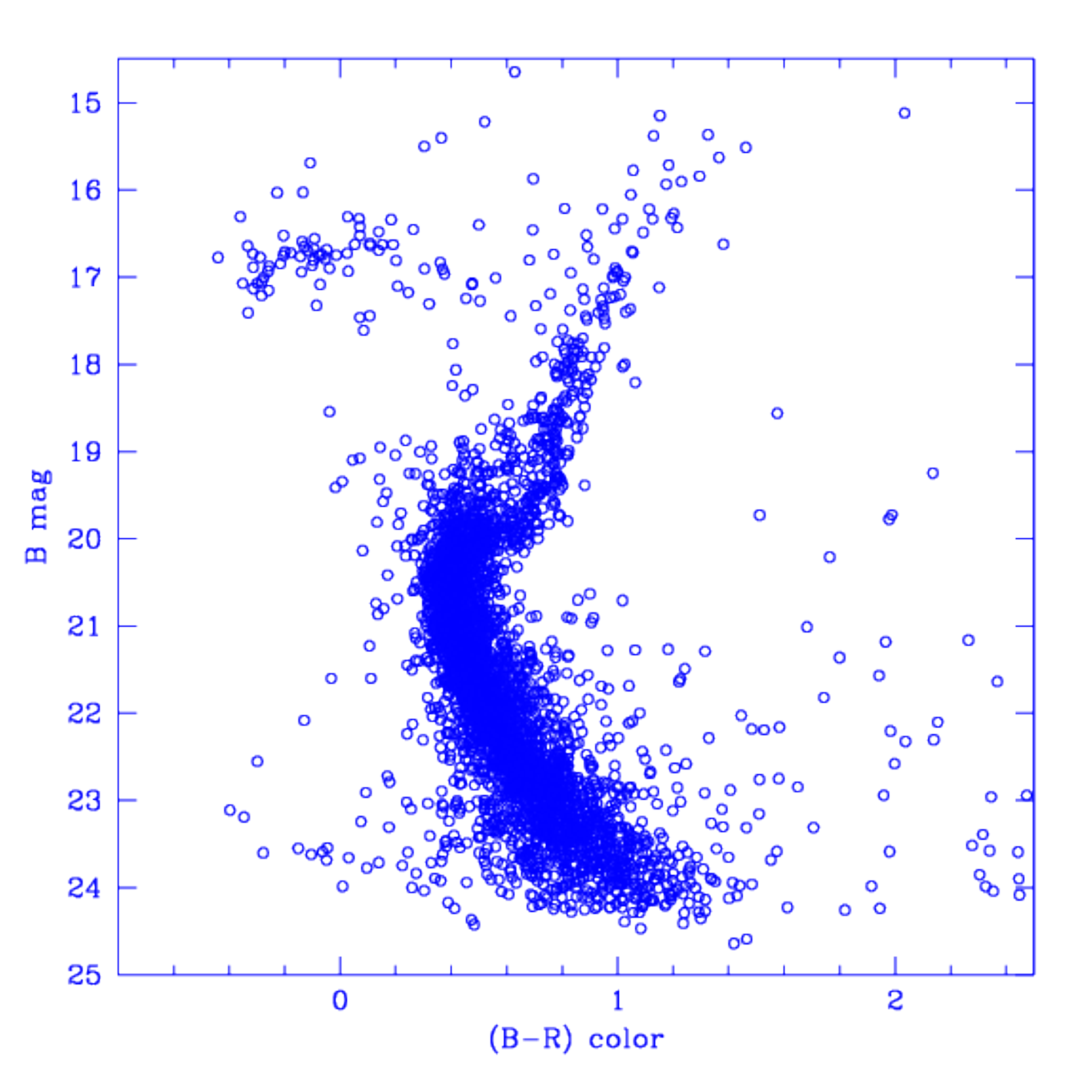}
	 \caption{
	The $B$ versus $(B-R)$ color-magnitude diagram (CMD) of the Globular cluster NGC 4147 as obtained using the 
	present calibration data taken using the 4K$\times$4K CCD mounted at the axial port of the 3.6m DOT. 
	The total number of common stars plotted (detected in both the filters) are around 3500 with a photometric 
	accuracy of $<$ 0.2 mag. In this figure, the number of detected stars having B $<$ 24 mag are around 150
	(with a photometric accuracy of $<$ 0.2 mag) in the effective exposure time of 1200 sec. There are many more detections 
	for which error could not be established using the present data set. The main sequence branch and other features typical 
	for this cluster are clearly identified using our data set observed in B and R filters.} 
	\end{minipage}
	\end{figure}

	\section{Calibrations and First light results}\label{sec:intro}

	On 23rd March 2017 images of Globular cluster NGC 4147 and several Landolt 
	standard fields were acquired both in open and closed loop conditions in various
	filters along with bias and flat frames. The range of exposure times were 30 sec to 600 sec
	in various filters to cover both bright and faint point sources. Using IRAF/DAOPHOT standard routines,
	circularity and radial profiles were checked and pre-processing was done (both NGC 4147 and Landolt PG 
	standard fields). The FWHM of stars are nearly similar within the single frame of observations. The range of 
	FWHM varies from frame to frame from 1.6 arcsec to 1.2 arcsec later in the night as seeing improved. 

	\noindent
	We also attempted calibrating Globular cluster field NGC 4147 and acquired data in B and R filters with 
	an exposure time of 60 sec to 600 sec to cover both bright and faint stars in the field. For the 
	calibration purposes, we also acquired Landolt standard field PG1525 and PG1528 in BVR filters having a 
	good color range around similar airmass values. 

	\noindent
	The pre-processing of the images were done using the bias and flat frames observed in respective filters 
	were used to pre-process the images using standard IRAF routine called ''ccdproc'' after trimming the 
	images to remove the over-scan area. We used ''cosmicrays'' routine to remove the cosmic hits in various 
	exposures and images were stacked wherever required. After the pre-processing is done, we used the 
	DAOPHOT-II FORTRAN subroutines in sequential standard orders (i.e. daogrow, ndaomatch, ndaomster, nccdstd, 
	nccdlib, nfinal etc.) to calculate the transformation coefficients determined using the observed Landolt 
	standard fields. These transformations were then applied to the observed globular cluster field NGC 4147 
	to calibrate the field. The calibrated color magnitude diagram (CMD) of the NGC 4147 (Figure 11) obtained 
	using this procedure was plotted in Figure 12 and results are described in the caption. 
	Atmospheric extinction coefficients in all 10 filters were also estimated by observing Landolt standard fields
	within airmass range of 1.1 to 3.5 and the values obtained using standard procedure were determined.
	The derived values of the extinction coefficients using the data taken during 16-17 April 2017 are listed in 
	the last column of Table 2.  
	%
	\section*{Acknowledgments}
	Authors thankfully acknowledge contributions from Er. Vishal Shukla,
	Dr. Amitesh Omar, Dr. Brijesh Kumar and the full technical team of the 
	DOT project. Timely support and encouragements from Dr. A. K. Pandey, 
	Dr. Wahab Uddin and Prof. Ram Sagar are acknowledged for their consistent support throughout and 
	to make this in-house instrument working for the astronomical community. It is suggested to cite this article
	while using data taken with the 4K$\times$4K CCD Imager.
	%
	%
	%

	\footnotesize
	\beginrefer

	\refer Alonso-Garcia J., Mateo M., Sen B. et al., 2012, AJ, 143, 70\\
	\refer Bekki K. \& Forbes D. A.,  2006, A\&A, 445, 485\\
	\refer Cameron P. B., Chandra P., Ray A. et al., 2004, Nature, 434, 1112\\
	\refer Deep A., Fiorentino G., Tolstoy E., et al., 2011, A\&A, 531, 151\\
	\refer Friel E. D., 1995, ARA\&A,  33, 381\\
	\refer Fabian A. C., 2012, ARA\&A, 50, 455\\
	\refer Howell S. B., 2006, Handbook of CCD Astronomy, second edition, ISBN 13978-0-521-85215-9\\
	\refer Joshi S., Mary D. L., Martinez P. et al., 2006, A\&A, 455, 303\\
	\refer Joshi Y. C., Pandey A. K., Narasimha D. \& Sagar R., 2005, A\&A, 433, 787\\
	\refer Kalirai J. S., Strader J., Anderson J., et al., 2008, ApJ, 682, 37\\
	\refer Kann D. A., Klose S., Zhang B. et al., 2011, ApJ, 734, 96\\
	\refer Kroupa P., 2002, Science, 295, 82\\
	\refer Kumar P. \& Zhang B., 2015, Physics Reports, 561, 1–109 \\
	\refer Kumar B., Sagar R., Rautela B. S. et al., Bull. Astr.  Soc. India, 28, 675 \\
	\refer Mayya Y. D., ``Photometric calibration of the CCD camera of 1-m telescope at VBO'', Journal of Astrophysics and Astronomy 12, 319\\
	\refer McLean I. S., 1989, in Electronic and Computer-Aided Astronomy, Ellis Horwood Chichester\\
	\refer Pandey A. K. \& Mahra H. S., 1986, Ap\&SS, 120, 107  \\
	\refer Pandey A. K., Durgapal, A. K.; Bhatt, B. C. et al., 1997, A\&AS, 122, 111  \\
	\refer Pandey A. K., Sharma, Saurabh; Ogura, K.; et al., 2008, MNRAS, 383, 1241  \\
	\refer Pandey J. C., Singh K. P.; Drake S. A. et al., 2005, Astronomical Journal, 130, 1231 \\
	\refer Pandey S. B., 2006, PhD Thesis  \\
	\refer Pandey S. B., 2013, JApA, 34, 157 \\
	\refer Pandey S. B., Sahu, D. K.; Resmi, L. et al., 2003, Bull. Astr.  Soc. India, 31, 19\\
	\refer Pant P, Stalin C. S. \& Sagar R., 1999, A\&A Suppl., 136, 19\\
	\refer Pasquini L. \& Belloni T., 1998, A\&A, 336, 902\\
	\refer Sagar, R., Kumar  B., Omar  A. \& Pandey  A. K., 2012, SPIE, 8444E, 1TS  \\
	\refer Sagar R. \& Pandey S. B., 2012, Astronomical Society of India Conference Series, 5, 1\\
	\refer Sagar, R., Subramaniam A., Richtler T. et al., 1999, A\&A Suppliment, 135, 391\\
	\refer Sagar  R., Piskunov  A. E., Miakutin  V. I., Joshi  U. C., 1986, MNRAS, 220, 383 \\
	\refer Samuel J. G. \& Ian R. S., 2011, Bull. Astr.  Soc. India, 39, 539\\
	\refer Saurabh S., Pandey A. K., Pandey J. C. et al., 2012, PASJ, 64, 107\\
	\refer Singal A. K., Konar C., Saikia D. J., 2004, MNRAS, 347, L79\\
	\refer Sreekumar P., 2005, Bull. Astr.  Soc. India, 33, 253\\
	\refer Stalin C. S., Sagar R. \& Pant P. et al., 2001, Bull. Astr.  Soc. India, 29, 39\\
	\refer Stalin C. S., Gopal-Krishna, Sagar, Ram \& Wiita Paul J., 2004, MNRAS, 350, 175\\
	\refer Swarup G., Ananthakrishnan S., Kapahi V. K. et al., 1991, Current Science, 60, 2 \\
	\refer William E. Harris \& Rene Racine, 1979, ARA\&A, 17, 241\\
	\refer Yadav R. K. S., Kumar B., Subramaniam A. et al., 2008, MNRAS, 390, 985\\

	\endrefer           
	\end{document}